\documentclass[12pt]{article}

\usepackage[utf8]{inputenc}
\usepackage[T1]{fontenc}
\usepackage{graphicx}
\usepackage{units}
\usepackage{dsfont}
\usepackage{upgreek}
\usepackage{amsfonts}
\usepackage{amsmath}
\usepackage{authblk}

\begin{document}

\title{Optical catastrophes of the swallowtail and butterfly beams}

\author[]{A. Zannotti}
\author[]{F. Diebel}
\author[]{M. Boguslawski}
\author[]{C. Denz}
\affil[]{Institute of Applied Physics and Center for Nonlinear Science (CeNoS), University of Münster, 48149 Münster, Germany}
\affil[]{\textit {a.zannotti@uni-muenster.de}}
\date{January 2017}
\maketitle

\begin{abstract}
We experimentally realize higher-order catastrophic structures in light fields to access the rich class of caustic swallowtail and butterfly beams. These beams present solutions of paraxial diffraction catastrophe integrals that are determined by potential functions, whose singular mapping manifests as caustic hypersurfaces in control parameter space. We systematically analyze the swallowtail and butterfly beams' caustics analytically and observe their field distributions experimentally in real and Fourier space. Their spectra can be described by polynomials or expressions with rational exponents capable to form a cusp.
\end{abstract}


\section{Introduction}

Caustics are ubiquitous in nature and our daily life, motivating scientists to investigate and utilize this general phenomenon in versatile fields, reaching from astrophysics~\cite{Diaferio1999, Nesvadba2006, Gifford2013} to models in social sciences~\cite{Oliva1981, Guastello1982, Sheridan1983, Hardy1996}, and oceanography~\cite{Chao1972, LeBlond1978, Mei1989}. Also, they are present in imaging of any kind of imperfect lens as well as in microscopic imaging~\cite{Kanaya1990, Patterson2008, Kennedy2011, Petersen2013, Tavabi2015}. For example, caustics play an important role in astrophysics, as the probability density function of particles (light, matter) distributed in a caustic network bears similarities to mass distribution in galaxies~\cite{Diaferio1999, Gifford2013} and may be related to predictions of the millennium simulation. Thus, understanding how these caustic networks form, their relation to extreme wave effects in the framework of freak and Rogue waves, as they occur, for instance, in oceans, is of high interest for a broad community of scientists in many fields of research~\cite{Hoehmann2010, Mathis2015}.  

In particular, the occurrence of caustics in light structures presents one of the oldest and most fundamental phenomena in optics~\cite{Born1970}. These incoherent ray-optic effects~\cite{Berry1975, Berry1979a} appear as high-intensity lines or surfaces. They are associated with supernumerary arcs close to rainbows~\cite{Berry1975} or may occur as bright lines of high intensity at the floor of shallow waters. Similar to their formation behind refractive lenses with imperfections, corresponding effects have been observed for numerous kinds of lenses with importance in optics, and surface analytics, in particular gravitational~\cite{Blandford1986, Wang1997} and electro-magnetic~\cite{Kennedy2011, Petersen2013, Tavabi2015} lenses. Overall, many fields will benefit from finding approaches to design high-contrast computational caustic mappings~\cite{Kiser2012, Ball2013, Schwartzburg2014}. 

Caustics in light are the manifestation of so-called catastrophes: Singularities in the gradient map of nonlinear potential functions that form as geometrically stable structures when perturbing external control parameters. The most prominent representative of catastrophe~\cite{Arnold2003} that manifests as caustic wave package is the fold catastrophe that was shown to be a solution of the quantum mechanical Schrödinger equation in form of the Airy function~\cite{Berry1979a}, and has been transferred and established in optics in 2007 as the paraxial Airy beam. One of the Airy beam's most exotic properties is the transverse invariant, accelerated movement of its caustic on a parabolic trajectory during propagation, yielding a wealth of applications~\cite{Baumgartl2008, Polynkin2009, Diebel2015}. The Airy beam thereby represents a spatial light structure besides many new light fields that have been explored as solutions of the paraxial Helmholtz equation in recent years capable to open new avenues in photonics. Among the most famous solutions we find generalizations of Gaussian beams~\cite{Woerdemann2011, Alpmann2015}, as well as nondiffracting beams~\cite{Rose2012}, like for instance Bessel beams~\cite{Durnin1987, Diebel2014}, as well as Mathieu and Weber beams~\cite{Diebel2016}, that were also designed to be self-accelerated~\cite{Belic2016}. However, the introduction of the fold catastrophe as paraxial Airy beam acts as a prelude for the renaissance to transfer higher-order catastrophes to optics. Subsequent to the Airy beam, in 2012 the Pearcey beam~\cite{Ring2012}, which represents the diffraction of a cusp catastrophe, emerges as a paraxial beam with intriguing features, like its form-invariance and auto-focusing propagation.

In general, the dimensionality of the investigated catastrophe causes caustics to occur as points, lines, surfaces or hypersurfaces, whose structure manifests as singular mapping of corresponding nonlinear potentials. Catastrophe theory predicts and analyses these potential functions and determines caustics as degenerated critical positions which show abrupt transitions of the observed system~\cite{Arnold2003}. Suggested by Arnol'd, a classification of seven elementary catastrophes determined by their potential functions has been theoretically established, corresponding to seven generic structures of bifurcation geometries~\cite{Zeeman1979}. These are, increasing in the order of catastrophe, given by: fold, cusp, swallowtail, butterfly, elliptical umbilic, hyperbolic umbilic, and parabolic umbilic~\cite{Arnold1975, Thom1975, Berry1980}. Although all of these catastrophes have been transferred by theoretical considerations to optical light structures~\cite{Berry1979, Berry1980, Nye2006}, only the creation the lowest-order catastrophes as Airy (fold) and Pearcey (cusp) beams have been experimentally realized in the paraxial approximation and utilized for applications. However, controlling caustics in all its parameters is essential for advanced, e.~g.~superresolution imaging and precision material processing technologies. Until now, they have not been artificially created beyond their occurrence in natural light fields.

Thus, with this work we present higher-order swallowtail and butterfly catastrophes as paraxial beams and systematically investigate the connection to their appearance and characteristics predicted by catastrophe theory. Therefore, we evaluate theoretically corresponding diffraction integrals and analyze their potential functions in order to find caustics as critical degenerated positions that strongly influence the beams' properties. Moreover, we demonstrate experimentally that their Fourier components are located on distributions that obey polynomial expressions.

We start with analyzing theoretically the class of caustic beams as well as their Fourier properties. Specifically, we contribute to analytical, numerical and experimental investigations with respect to the swallowtail and butterfly catastrophes by creating optical swallowtail and butterfly beams. Thereby, we describe their caustics with respect to control parameter space that defines the beam's properties. 

\section{Embedding the swallowtail catastrophe in paraxial beams}

We consider the \textit{caustic beam} $C_n(\mathbf{a})$ depending on one state variable $s$ emerging from the canonical diffraction catastrophe integral

\begin{equation}
C_n(\mathbf{a}) = \int\limits_{\mathds{R}}{e^{\text{i}P_n\left(\mathbf{a},s\right)}}\mathrm{d}s,
\label{eq:GeneralcausticBeam}
\end{equation}
whose properties are completely determined by the properties of the canonical potential function $P_n$ in the oscillating integrand

\begin{equation}
P_n\left(\mathbf{a},s\right) = s^n + \sum\limits_{j=1}^{n-2} \frac{a_j}{a_{0j}}s^j,
\label{eq:PolynomExponential}
\end{equation}
where $\text{dim}\left(\mathbf{a}\right) = n - 2 = d$. The potential function depends on a \textit{state variable} $s$ and \textit{control parameters} $a_j$~\cite{Berry1979}, which are components of the control parameter vector $\mathbf{a}$ with corresponding scaling factors $a_{0j} \in \mathds{R}^+$ to make the exponent dimensionless. $\mathbf{a}$ spans the control parameter space.

Caustics are defined as abrupt transitions of the optical ray system equivalent to bifurcations in nonlinear science by changing the number of crossing rays in each point in space~\cite{Berry1980}. Depending on the dimensionality $d$ of the control parameter space, the caustic arises as a point, line, surface or hypersurface. At these transitions, $P$ is stationary in $s$ for fixed $\mathbf{a}$. The caustics exist at points where the gradient mapping becomes singular, i.~e.~that the solutions of stationary $P$ are degenerated. Therefore, also the Hessian determinant of $P$ in $s$ has to vanish~\cite{Berry1979, Berry1980}. That is, we calculate caustic structures corresponding to

\begin{equation}
\frac{\partial P_n\left(\mathbf{a},s\right)}{\partial s} = 0, \quad \text{and } \quad  \frac{\partial^2 P_n\left(\mathbf{a},s\right)}{\partial s^2} = 0.
\label{eq:causticPositions}
\end{equation}

Note that the well known Airy $(n = 3)$ and Pearcey $(n = 4)$ beams emerge as special cases of $\left(\ref{eq:GeneralcausticBeam}\right)$ since $\text{Ai}(x) = C_3\left(x\right)$ and $\text{Pe}(x,y) = C_4\left(x, y\right)$. In the case of the Pearcey beam, we are able to identify the control parameters $\left(a_1, a_2\right)^T$ with the transverse spatial coordinates $\left(x, y\right)^T$ to directly obtain the transverse electric field distribution~\cite{Ring2012}. Instead, the famous 2D Airy beam is constructed by multiplying separable, orthogonal coordinates: $E = \text{Ai}(x) \cdot \text{Ai}(y)$~\cite{Siviloglou2007, Siviloglou2007a} and therefore does not present a generic structure predicted by catastrophe theory. 

In the following, we develop a construction scheme to create higher-order ($n \geq 5$) caustic beams by identifying two control parameters $(a_\alpha, a_\beta)^T$ with spatial transverse coordinates $(x, y)^T$, where $\alpha, \beta \in \mathds{N}^+$ and $1 \leq \alpha, \beta \leq n$. The remaining control parameters are chosen as being constant and play an important role in controlling further properties of the realized beams.

Swallowtail beams result as solutions $\text{Sw}(x,y) = C_5\left(x, y, \mathbf{a}'\right)$, where $\mathbf{a}'$ is one of the control parameters that was chosen to be constant. Thus, \textbf{three} different swallowtail beams are can be constructed, from which each exhibits unique features. However, all share characteristic properties of a swallowtail catastrophe. Furthermore, we create and analyze butterfly beams by calculating $\text{Bu}(x,y) = C_6\left(x, y, \mathbf{a}'\right)$. Here, $\mathbf{a}'$ are two control parameters that stay constant. Consequently, \textbf{six} different butterfly beams complete the class of butterfly beams. All of them exhibit characteristic properties determined by the butterfly catastrophe. 

We start our investigations on the swallowtail beams $\text{Sw}(\mathbf{a})$ demonstrating its of fundamental transverse field distributions, and subsequently connect it with the control parameter space.

\begin{figure}
\centering
\includegraphics[width=.7\columnwidth]{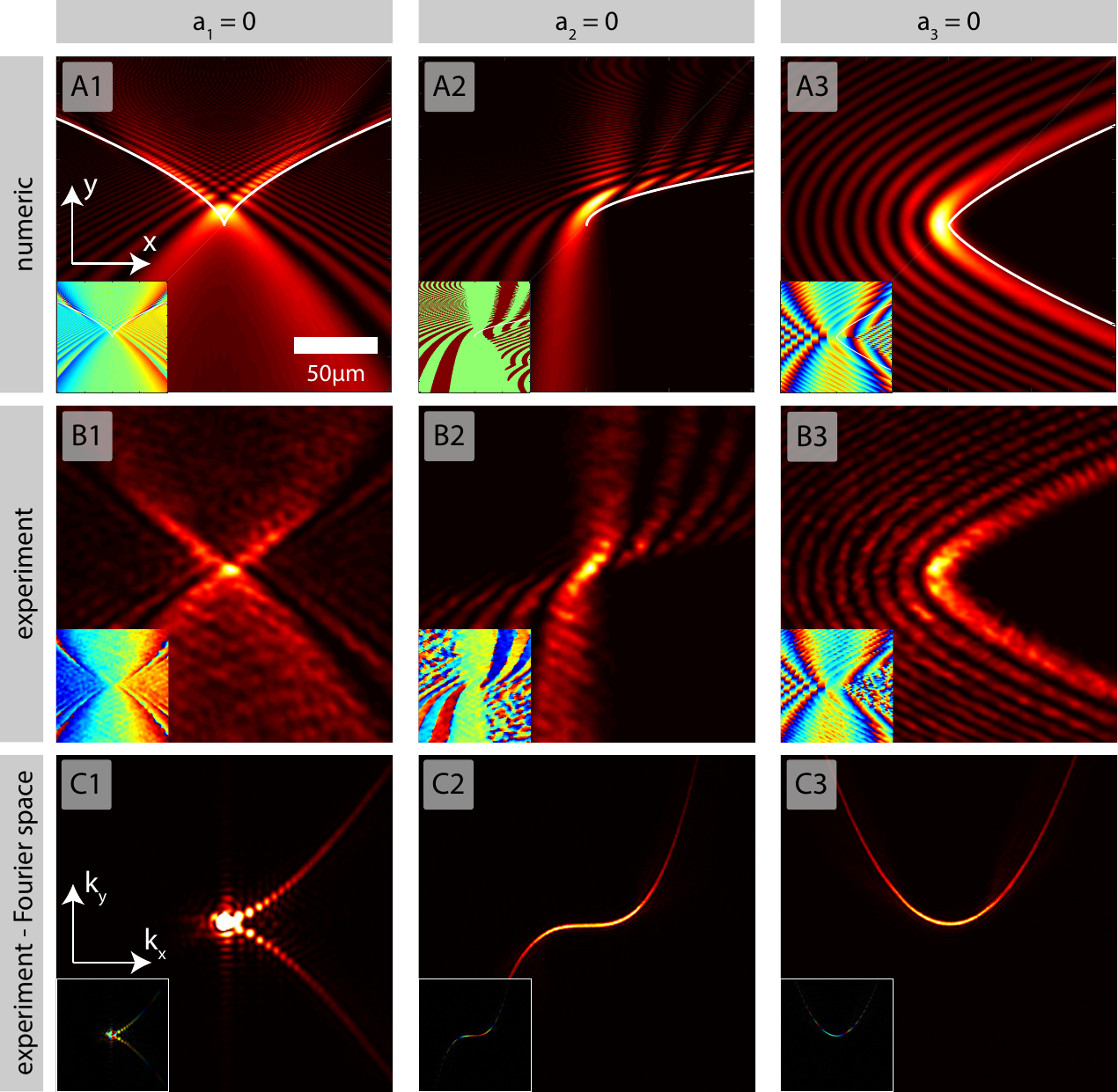}%
\caption{Three swallowtail beams arise by setting always one other control parameters constant. Numerically calculated light fields (A) and their respective caustics (white lines) are shown. Experimentally, the lights fields were measured in real (B) and Fourier space (C).}%
\label{fig:Sw3Beams}%
\end{figure}

The three swallowtail beams of this case of artificially created and tailored caustic beams are shown in their most fundamental appearance in Fig. \ref{fig:Sw3Beams}. In each case we set the constant control parameter to zero. The beams are constructed by numerically solving equation $\eqref{eq:GeneralcausticBeam}$ using adoptions of methods as described in~\cite{Connor1982}. These numerical results are shown in A, and were as well realized in experiments, shown in B $\&$ C, where beneath the real space distributions of B, Fourier spectra~C have additionally been recorded. Insets show corresponding calculated and measured spatial phase distributions. Light fields have been created using a phase-only spatial light modulator (SLM) that addresses light fields in real space. The transverse scale is chosen to be $\unit[200 \times 200]{\upmu m}$, and characteristic scale factors were $a_{01} = a_{02} = a_{03} = \unit[50]{\upmu m}$. Experimental details are given in the SM.

White lines in A are calculated solutions of equation~$\left(\ref{eq:causticPositions}\right)$ and therefore represent cross sections through the caustic surface in parameter space $\mathbf{a}$. We performed a parametrization of the caustic surface in dependence of one parametrization parameter $u$ and the remaining constant control parameter. The resulting expressions are given in the SM.

A 2D Fourier transform of Eq. \eqref{eq:GeneralcausticBeam} for the swallowtail ($n = 5$) and butterfly beams ($n = 6$) are performed analytically. For the explicit derivation of the Fourier transforms of these caustic beams we refer to the Supplemental Material (SM).

We pronounce the simplicity of the Fourier spectrum of caustic beams, since these are located on polynomial expressions of degree up to $n - 2$ or expressions with rational exponents that are capable of forming cusps. For instance, the Pearcey beam's Fourier components are located on a parabola with $\exp\left[\text{i} k_x^4\right]$ phase distribution~\cite{Ring2012}. In the SM we show that the $\text{Sw}(x,y,a_3)$ beam (C3) exhibits a parabolic $\left(k_y = k_x^2\right)$, the $\text{Sw}(x,a_2,y)$ beam (C2) a cubic $\left(k_y = k_x^3\right)$, and the $\text{Sw}(a_1,x,y)$ beam (C1) even a cuspoid $\left(k_y^2 = k_x^3\right)$ spectral distribution with corresponding phase functions, depending on which cross section in control parameter space $\mathbf{a}$ (i.e. choice of $a_\alpha,~a_\beta$) is regarded. Note that by increasing the order $n$ of a caustic beam, new distributions of spectral components do not always arise. A good example is the swallowtail beam where $a_3 = $const. whose Fourier components were located on a parabola similar to those of a Pearcey beam. Similarly, we show in the SM that butterfly beams have, beneath parabolic and cubic distributions, additionally quartic and cuspoid $\left(k_y^3 = k_x^4\right)$ Fourier component distributions.

Catastrophes form geometrically stable structures where the bifurcation characteristics of the nonlinear potential function suddenly change. The transfer to optics leads to caustics that mark separated areas of different numbers of crossing light rays. Here we clearly see that the $\text{Sw}(x,y,0)$ beam in A3 exhibits an area of two-beam interference with spatially varying $\mathbf{k}$-vectors on the left side of the caustic and performs an abrupt transition at the caustic interface to zero beams on the right side of low intensity. 

Numerically calculated swallowtail beams are in very good agreement with the experimental realizations in both, spatial intensity and phase distributions. The predicted caustics indicated as white lines have been parametrized according to equation $\left(\ref{eq:causticPositions}\right)$ and fit very well the numerical realizations in scale and form. One of the most striking characteristics of the beams, strongly determining their properties, is the distribution of Fourier coefficients, which is perfectly predicted by the theoretical description stated in the SM.

Note that we intentionally omit any discussion of apodizing exponential or Gaussian functions in order to keep the total intensity of the light fields finite, as was suggested for both, the Airy~\cite{Siviloglou2007} and Pearcey beam~\cite{Ring2012}. The fundamental physics of the infinite power beams discussed here resembles that of the apodized ones, and analytic expressions would become extensively large. Nevertheless, for the experimental realization the finite transverse dimensions of the light fields are determined by the size of the SLM.

\begin{figure}
\centering
\includegraphics[width=\columnwidth]{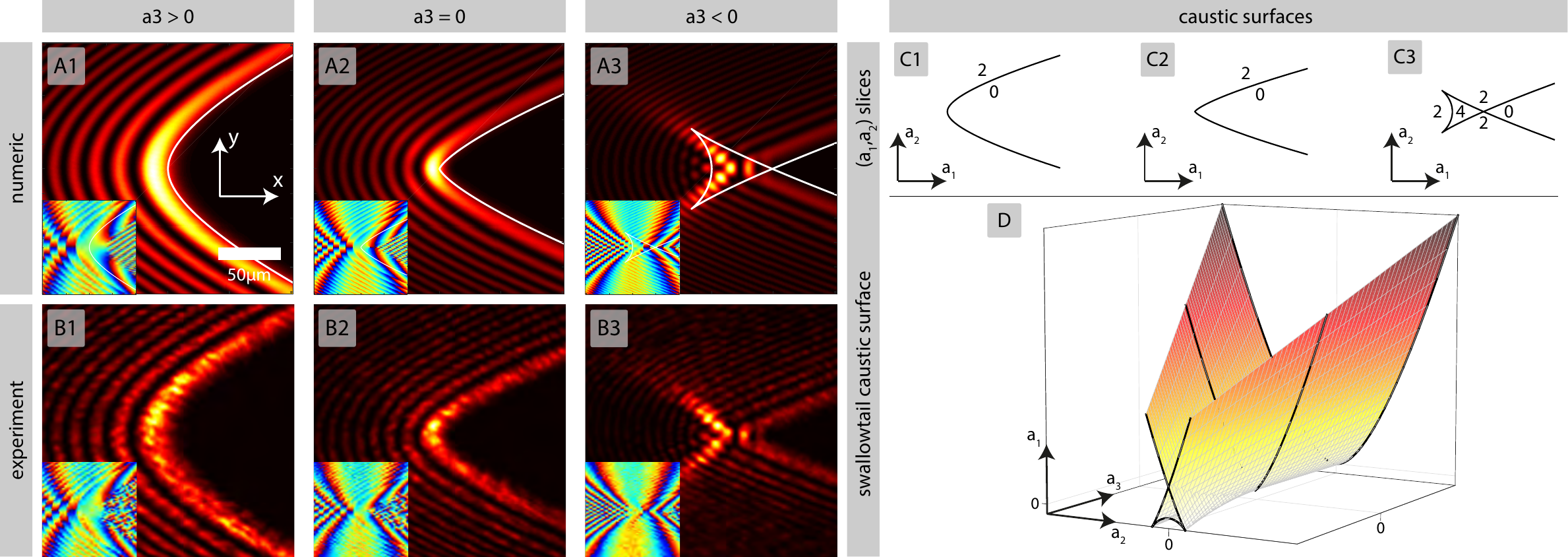}%
\caption{We resemble the swallowtail catastrophe's most striking characteristic at the corresponding swallowtail beams: Fingerprints of the caustic surface manifested in light. Shown are realizations of $\text{Sw}(x,y,a_3)$ beams in simulations (A) and experiment (B) for different parameters of $a_3$. In the $(x, y)$ plane, the for $(a_3 > 0)$ present fold catastrophe abruptly changes at the cusp point $(a_3 = 0)$ into two cusps, that encircle a region of 4 beam interference. Images C visualize characteristic cross sections through the caustic surface of the swallowtail catastrophe which is shown in D.}%
\label{fig:SwDifferentA3Values}%
\end{figure}

\section{Cross sections of the swallowtail caustic}

Each caustic beam's properties are connected to the distribution and kind of their correspondent caustics, may they consist of fold, cusp, or a combination of these catastrophes, which can be found by analyzing the potential functions. In order to investigate and image the caustic surface of the swallowtail catastrophe of control parameter space in different light structures, we arbitrarily chose $a_3$ to be constant and create $\text{Sw}(x,y,a_3)$ beams with different values of $a_3$. Thus, by varying $a_3$ a sliced scan of the surface in control parameter space can be performed whose properties manifest at the light fields.

The resulting swallowtail beams are depicted in Fig.~\ref{fig:SwDifferentA3Values} A $\&$ B with varying control parameter $a_3$, which performs a transition from positive (A1) to zero (A2) to negative (A3) values. This allows us to trace and illustrate fundamental properties of the swallowtail catastrophe that now manifests in the corresponding light structure: the caustic surface in control parameter space $\mathbf{a}$ is defined by abrupt transitions of the system between areas of different numbers of crossing beams (in the regime of ray optics)~\cite{Kravtsov1983}. By varying the control parameter $a_3$, we have demonstrated the transition of the system from a swallowtail beam that consists of areas with 2 beam interference and areas with 0 beams ($a_3 > 0$) over the cuspoid point at $a_3 = 0$ to a swallowtail beam that clearly shows areas of 2 as well as 4 beams interference and low intensity areas (0 beams) for $a_3 < 0$, as depicted schematically in C. These system transitions of the swallowtail catastrophe are additionally imaged by the caustic surface in control parameter space in Fig.~\ref{fig:SwDifferentA3Values}D, which is in perfect agreement with previous discussions~\cite{Arnold2003, Nye2006, Berry1980}. The caustic surface results from the parametrization for constant $a_3$ described in the SM.

\section{The butterfly caustic}

Each beam of the class of caustic beams that depends on one state variable $s$, e.g. Airy, Pearcey or the three swallowtail beams, is striking due to its unique properties. Their static intensity and phase distributions in real and Fourier space are closely connected to catastrophe theory. 

Similarities in the properties of these different beams manifest predominantly in their Fourier spectra. Discussions in the SM show that by increasing the order $n$ of the caustic beam, new properties will always emerge, since completely new distributions of Fourier components, like for instance higher-order polynomials, will arise. 

Therefore, we introduce the six butterfly beams that complete the transfer from catastrophe theory to optics for all catastrophes that belong to the elementary catastrophes and are determined by one state variable $s$. Those butterfly beams, whose Fourier spectra cannot be expressed in terms of lower-order caustic beams, are worth investigating thoroughly, since these promising light fields will show new individual propagation characteristics.

\begin{figure}
\centering
\includegraphics[width=1\columnwidth]{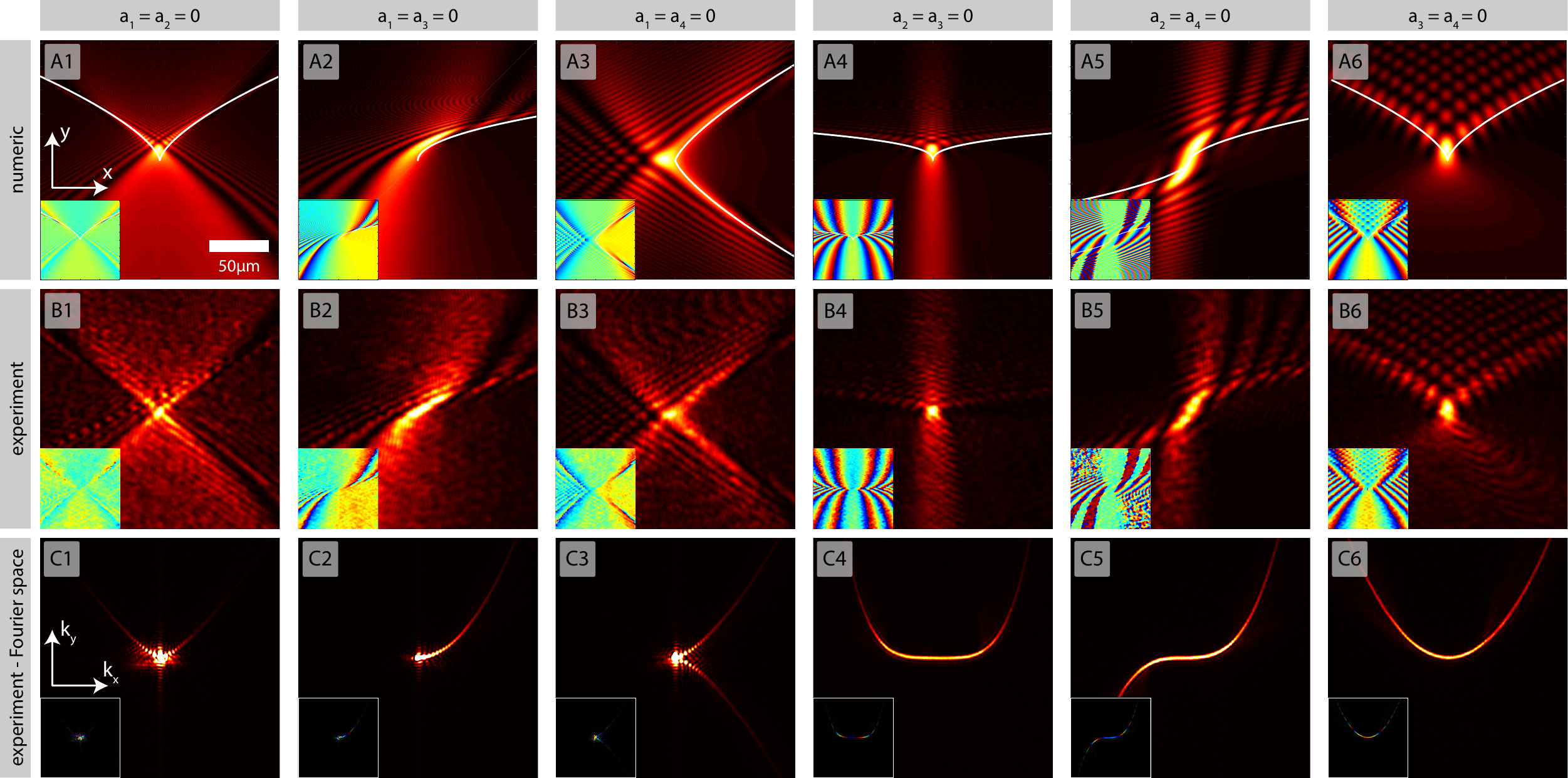}%
\caption{Butterfly beams produced by setting various control parameters constant, imaged in intensity and phase. Real-space numerical simulations (A) are supported by experimental realizations (B). Additionally, experimentally obtained Fourier spectra are shown (C).}%
\label{fig:butterflyBeam}%
\end{figure}

Fig. \ref{fig:butterflyBeam} shows the transverse light field distributions for each of the six butterfly beams in both, real and Fourier space. Numerical calculations of equation \eqref{eq:GeneralcausticBeam} are in very good agreement with the experimental realizations. Insets show corresponding calculated or measured spatial phases. Again, the experimentally obtained Fourier spectra are represented by polynomials and expressions with rational exponents, as predicted in the SM.

Similar to the treatment of the swallowtail catastrophes, we state a parametrization for the butterfly catastrophe in the SM. The parametrization was performed for one parameter $u$ and the respective two constant control parameters. For this higher-order catastrophe, control parameter space is four-dimensional. We restricted ourselves to image caustic lines (indicated in Fig.~\ref{fig:butterflyBeam}A as white lines) by choosing corresponding cross sections in parameter space. Areas with different numbers of crossing rays are clearly observable.

For constant $a_1$ the intensity of the spectra always diverges at the origin of Fourier space, which was also observed experimentally where the intensity increases extremely (but finitely). Again, the transverse picture section is chosen to be $\unit[100 \times 100]{\upmu m}$, and characteristic scale factors were $a_{01} = a_{02} = a_{03} = a_{04} = \unit[50]{\upmu m}$. All constant control parameters are zero.

\section{Conclusion}

To conclude, we extended the diversity of caustic structures in paraxial light by numerically and experimentally embedding swallowtail and butterfly caustics in artificially tailored transverse light structures. In the framework of catastrophe theory we proved that the key properties of the related catastrophes are preserved in the paraxial regime. Thus, we demonstrated that emerging characteristics of higher-order swallowtail and butterfly catastrophes manifest at these optical paraxial beams in terms of abrupt transitions of the number of crossing beams. Dynamics of the potential function of the diffraction integral, which highly influences the beams' properties, were investigated with respect to control parameter space. Thereby, we showed cross sections through three- and four-dimensional control parameter space and shaped the beams according to corresponding parametrized caustic (hyper-) surfaces.

Fourier spectra of these intriguing beams were calculated analytically and observed experimentally. We demonstrated, that spectral components of these caustics beams are distributed on polynomial expressions of increasing degree or are located on expressions with rational exponents. Some spectra are calculated to exhibit diverging energies at the origin, resulting in highly increasing but finite energies in experimental realizations. 

By accessing higher-order catastrophic light structures, our findings allow investigating fundamental characteristics and properties of catastrophes in optics and in particular to utilize the caustics' high intensities and propagation properties beyond the established caustic Airy and Pearcey beams. The trajectories and focusing effects of caustic light are promising to be discussed for designing tailored waveguides, particle manipulation, material processing, or improved super imaging in microscopic and sub-diffractive applications.

\section*{Acknowledgement}

We gratefully acknowledge fruitful discussions with Sir Michael Berry and Mark Dennis.


\begin{thebibliography}{10}
\expandafter\ifx\csname url\endcsname\relax
  \def\url#1{{\tt #1}}\fi
\expandafter\ifx\csname urlprefix\endcsname\relax\def\urlprefix{URL }\fi
\providecommand{\eprint}[2][]{\url{#2}}

\bibitem{Diaferio1999}
Diaferio A 1999 {\em Mon. Not. R. Astron. Soc.\/} {\bf 309} 610--622

\bibitem{Nesvadba2006}
Nesvadba N, Lehnert M, Eisenhauer F, Genzel R, Seitz S, Davies R, Saglia R,
  Lutz D, Tacconi L, Bender R and Abuter R 2006 {\em The Astrophysical
  Journal\/} {\bf 650} 661--668

\bibitem{Gifford2013}
Gifford D, Miller C and Kern N 2013 {\em The Astrophysical Journal\/} {\bf 773}
  116

\bibitem{Oliva1981}
Oliva T~A, Peters M~H and Murthy H~S~K 1981 {\em Behavioural Science\/} {\bf
  26} 153--162

\bibitem{Guastello1982}
Guastello S~J 1982 {\em Behav. Sci.\/} {\bf 27} 259--272

\bibitem{Sheridan1983}
Sheridan J~E and Abelson M~A 1983 {\em AMJ\/} {\bf 26} 418--436

\bibitem{Hardy1996}
Hardy L 1996 {\em The Sport Psychologist\/} {\bf 10} 140--156

\bibitem{Chao1972}
Chao Y~Y and Pierson W~J 1972 {\em Journal of Geophysical Research\/} {\bf 77}
  4545 -- 4554

\bibitem{LeBlond1978}
LeBlond P~H and Mysak L~A 1978 {\em Waves in the Ocean\/} (Elsevier
  Oceanography Series) chap Waves near a caustic

\bibitem{Mei1989}
Mei C~C 1989 {\em Applied Dynamics of Ocean Surface Waves (Advanced Series on
  Ocean Engineering, V. 1)\/} (World Scientific Publishing Co Pte Ltd) chap The
  Neighborhood of a Straight Caustic

\bibitem{Kanaya1990}
Kanaya K, Oho E, Adachi K, Yamamoto Y and Doi H 1990 {\em Micron And
  Microscopica Acta\/} {\bf 21} 57--68

\bibitem{Patterson2008}
Patterson E~A and Whelan M~P 2008 {\em Nanotechnology\/} {\bf 19} 105502

\bibitem{Kennedy2011}
Kennedy S~M, Zheng C~X, Tang W~X, Paganin D~M and Jesson D~E 2011 {\em
  Ultramicroscopy\/} {\bf 111} 356--363

\bibitem{Petersen2013}
Petersen T~C, Weyland M, Paganin D~M, Simula T~P, Eastwood S~A and Morgan M~J
  2013 {\em Phys. Rev. Lett.\/} {\bf 110} 1--5

\bibitem{Tavabi2015}
Tavabi A~H, Migunov V, Dwyer C, Dunin-Borkowski R~E and Pozzi G 2015 {\em
  Ultramicroscopy\/} {\bf 157} 57--64

\bibitem{Hoehmann2010}
Höhmann R, Kuhl U, Stöckmann H~J, Kaplan L and Heller E~J 2010 {\em {Phys.
  Rev. Lett.}\/} {\bf 104} 093901

\bibitem{Mathis2015}
Mathis A, Froehly L, Toenger S, Dias F, Genty G and Dudley J~M 2015 {\em
  Scientific Reports\/} {\bf 5} 12822

\bibitem{Born1970}
Born M and Wolf E 1970 {\em {Principles of Optics}\/} {4th} ed ({Pergamon
  Press})

\bibitem{Berry1975}
Berry M~V 1975 {\em J. Phys. A: Math. Gen.\/} {\bf 8} 566--584

\bibitem{Berry1979a}
Berry M~V and Balazs N~L 1979 {\em Am. J. Phys.\/} {\bf 47} 264--267

\bibitem{Blandford1986}
Blandford R and Narayan R 1986 {\em The Astrophysical Journal\/}  568--582

\bibitem{Wang1997}
Wang Y and Turner E 1997 {\em Mon. Not. R. Astron. Soc.\/} {\bf 292} 863--870

\bibitem{Kiser2012}
Kiser T and Pauly M 2012 {\em {Caustic Art}\/} (EPFL Technical Report)

\bibitem{Ball2013}
Ball P 2013 {\em New Scientist\/} {\bf 217} 40--43

\bibitem{Schwartzburg2014}
Schwartzburg Y, Testuz R, Tagliasacchi A and Pauly M 2014 {\em ACM Trans.
  Graph.\/} {\bf 33} 74

\bibitem{Arnold2003}
Arnol'd V~I and Wassermann G~S 2003 {\em {Catastrophe Theory}\/} {3rd} ed
  ({Springer Verlag})

\bibitem{Baumgartl2008}
Baumgartl J, Mazilu M and Dholakia K 2008 {\em Nature Photon.\/} {\bf 2}
  675--678

\bibitem{Polynkin2009}
Polynkin P, Kolesik M, Moloney J~V, Siviloglou G~A and Christodoulides D~N 2009
  {\em Science\/} {\bf 324} 229--232

\bibitem{Diebel2015}
Diebel F, Bokic B~M, Timotijevic D~V, {Jovic Savic} D~M and Denz C 2015 {\em
  Opt. Express\/} {\bf 23} 24351

\bibitem{Woerdemann2011}
Woerdemann M, Alpmann C and Denz C 2011 {\em Appl. Phys. Lett.\/} {\bf 98}
  111101

\bibitem{Alpmann2015}
Alpmann C, Schöler C and Denz C 2015 {\em Appl. Phys. Lett.\/} {\bf 106}
  241102

\bibitem{Rose2012}
Rose P, Boguslawski M and Denz C 2012 {\em New J. Phys.\/} {\bf 14} 033018

\bibitem{Durnin1987}
Durnin J, Miceli J~J and Eberly J~H 1987 {\em Phys. Rev. Lett.\/} {\bf 58}
  1499--1501

\bibitem{Diebel2014}
Diebel F, Rose P, Boguslawski M and Denz C 2014 {\em Appl. Phys. Lett.\/} {\bf
  104} 191101

\bibitem{Diebel2016}
Diebel F, Rose P, Boguslawski M and Denz C 2016 {\em New J. Phys.\/} {\bf 18}
  053038

\bibitem{Belic2016}
Zhang Y, Liu J, Wen F, Li C, Zhang Z, Zhang Y and Belic M~R 2016 {\em arXiv\/}
  {\bf 1605.09525v1}

\bibitem{Ring2012}
Ring J~D, Lindberg J, Mourka A, Mazilu M, Dholakia K and Dennis M~R 2012 {\em
  Optics Express\/} {\bf 20} 18955--18966

\bibitem{Zeeman1979}
Zeeman E~C 1979 {\em Structural Stability in Physics\/} (Berlin, Heidelberg:
  Springer Berlin Heidelberg) chap Catastrophe Theory, pp 12--22

\bibitem{Arnold1975}
Arnol'd V~I 1975 {\em Russ. Math. Survs.\/} {\bf 30} 1--75

\bibitem{Thom1975}
Thom R 1975 {\em Trans. by D. Fowler. Reading, Mass.: Benjamin\/}

\bibitem{Berry1980}
Berry M~V and Upstill C 1980 {\em Prog. in Optics\/} {\bf 28} 257--346

\bibitem{Berry1979}
Berry M~V, Nye J~F and Wright F~J 1979 {\em Phil. Trans. R. Soc. Lond. A\/}
  {\bf 291} 453--484

\bibitem{Nye2006}
Nye J 2006 {\em Proceedings of the Royal Society A: Mathematical, Physical and
  Engineering Sciences\/} {\bf 462} 2299--2313

\bibitem{Siviloglou2007}
Siviloglou G~A and Christodoulides D~N 2007 {\em Opt. Lett.\/} {\bf 32}
  979--981

\bibitem{Siviloglou2007a}
Siviloglou G~A, Broky J, Dogariu A and Christodoulides D~N 2007 {\em Phys. Rev.
  Lett.\/} {\bf 99} 213901

\bibitem{Connor1982}
Connor J~N~L and Curtis P~R 1982 {\em J. Phys. A: Math. Gen.\/} {\bf 15}
  1179--1190

\bibitem{Kravtsov1983}
Kravtsov Y~A and Orlov Y~I 1983 {\em Usp. Fiz. Nauk\/} {\bf 141} 591--627

\end{thebibliography}
\providecommand{\newblock}{}

\newpage

\setcounter{equation}{0}
\setcounter{page}{1}

\begin{center}
\section*{Supplementary Material | \small Optical catastrophes of the swallowtail and butterfly beams \normalsize}
\end{center}

\subsection*{Fourier spectra of swallowtail and butterfly beams}

We perform a 2D Fourier transform of equation (1) in the main article regarding two control parameters $a_\alpha,~a_\beta$. Corresponding Fourier frequencies of $\mathbf{a}$ with components $a_j, j = 1,...,n$ are $\mathbf{k}$ with components $k_j$, i.e. here we introduce $k_\alpha, k_\beta$. With $\mathbf{a}'$ we denote all remaining control parameters $\mathbf{a}$ except $a_\alpha,~a_\beta$. We neglected $2\pi$ scaling factors for reasons of clarity.

We identify two control parameters $(a_\alpha,~a_\beta)^T$ with spatial transverse coordinates $(x, y)^T$ and treat remaining control parameters $\mathbf{a}'$ as constants. Thus, in case of the swallowtail beam $\text{Sw}$ $(n = 5)$, which is defined by three control parameters $(a_1,~a_2,~a_3)^T$, three different transverse field distributions can be created by keeping constant one control parameter. All resulting fields show characteristics of the swallowtail catastrophe.

\setlength\jot{18pt}
\begin{eqnarray}
\nonumber
\tilde{\text{Sw}}\left(a_1,k_x,k_y\right) &= &\delta_{k_x,r^+} \cdot \frac{x_{0}y_{0}}{\left|2\left(x_{0}k_x\right)^\frac{1}{2}\right|}\cdot \left[ \delta\left(\left(x_{0} k_x\right)^\frac{3}{2} - y_{0}k_y\right) \cdot e^{\text{i}\left(x_{0}k_x\right)^\frac{5}{2}}e^{\text{i}\frac{a_1}{a_{01}}\left(x_{0}k_x\right)^\frac{1}{2}} \right. \\ 
&+ &\left. \delta\left(-\left(x_{0} k_x\right)^\frac{3}{2} - y_{0}k_y\right) \cdot e^{-\text{i}\left(x_{0}k_x\right)^\frac{5}{2}}e^{-\text{i}\frac{a_1}{a_{01}}\left(x_{0}k_x\right)^\frac{1}{2}} \right]\\
\tilde{\text{Sw}}\left(k_x,a_2,k_y\right) &= &x_{0}y_{0} \cdot \delta\left(\left(x_{0} k_x\right)^3 - y_{0}k_y\right) \cdot e^{\text{i}\left(x_{0}k_x\right)^5} e^{\text{i}\frac{a_2}{a_{02}}\left(x_{0}k_x\right)^2}\\
\tilde{\text{Sw}}\left(k_x,k_y,a_3\right) &= &x_{0}y_{0} \cdot \delta\left(\left(x_{0} k_x\right)^2 - y_{0}k_y\right) \cdot e^{\text{i}\left(x_{0}k_x\right)^5} e^{\text{i}\frac{a_3}{a_{03}}\left(x_{0}k_x\right)^3}
\end{eqnarray}

Thereby, $r^+ \in \mathds{R}_0^+$. The Kronecker delta $\delta_{i,j}$ equals $1$ if $i = j$ and equals $0$ for $i \neq j$. Thus, $\delta_{k_x,r^+}$ only allows positive $k_x$ (including $k_x = 0$) to contribute in the corresponding term. The spectra for the beams, as calculated above, exhibit Fourier components that are each only located in two quadrants: For $a_1 = $const., the spectrum is restricted to $k_x \geq 0$. For $a_2 = $const., Fourier components are only allowed for $k_x \geq 0 \vee k_y \geq 0$ and $k_x \leq 0 \vee k_y \leq 0$, whereas for $a_3 = $const., the spectrum is restricted to $k_y \geq 0$. 

Expanding our approach to butterfly beams $\text{Bu}$ $(n = 6)$ and still identifying two control parameters with spatial coordinates, in this case six transverse field distributions can be created.

\setlength\jot{24pt}
\begin{eqnarray}
\tilde{\text{Bu}}\left(a_1,a_2,k_x,k_y\right) &= &\frac{x_{0}y_{0}}{\left|3\left(x_{0}k_x\right)^\frac{2}{3}\right|} \cdot \delta\left(\left(x_{0} k_x\right)^\frac{4}{3} - y_{0}k_y\right) \cdot e^{\text{i}\left(x_{0}k_x\right)^2}e^{\text{i}\frac{a_1}{a_{01}}\left(x_{0}k_x\right)^\frac{1}{3}}e^{\text{i}\frac{a_2}{a_{02}}\left(x_{0}k_x\right)^\frac{2}{3}}\\
\nonumber
\tilde{\text{Bu}}\left(a_1,k_x,a_3,k_y\right) &= &\delta\left(\left(x_{0} k_x\right)^2 - y_{0}k_y\right) \cdot \delta_{k_x,r^+} \cdot \frac{x_{0}y_{0}}{\left|2\left(x_{0}k_x\right)^\frac{1}{2}\right|} \cdot e^{\text{i}\left(x_{0}k_x\right)^3} \\ &\cdot &\left[ e^{\text{i}\frac{a_1}{a_{01}}\left(x_{0}k_x\right)^\frac{1}{2}}e^{\text{i}\frac{a_3}{a_{03}}\left(x_{0}k_x\right)^\frac{3}{2}} + e^{-\text{i}\frac{a_1}{a_{01}}\left(x_{0}k_x\right)^\frac{1}{2}}e^{-\text{i}\frac{a_3}{a_{03}}\left(x_{0}k_x\right)^\frac{3}{2}} \right]\\
\nonumber
\tilde{\text{Bu}}\left(a_1,k_x,k_y,a_4\right) &= &\frac{x_{0}y_{0}}{\left|2\left(x_{0}k_x\right)^\frac{1}{2}\right|}  \cdot \delta_{k_x,r^+} \cdot \left[ \delta\left(\left(x_{0} k_x\right)^\frac{3}{2} - y_{0}k_y\right) \cdot e^{\text{i}\left(x_{0}k_x\right)^3}e^{\text{i}\frac{a_1}{a_{01}}\left(x_{0}k_x\right)^\frac{1}{2}}e^{\text{i}\frac{a_4}{a_{04}}\left(x_{0}k_x\right)^2} \right. \\
&+ &\left. \delta\left(-\left(x_{0} k_x\right)^\frac{3}{2} - y_{0}k_y\right) \cdot e^{\text{i}\left(x_{0}k_x\right)^3}e^{-\text{i}\frac{a_1}{a_{01}}\left(x_{0}k_x\right)^\frac{1}{2}}e^{\text{i}\frac{a_4}{a_{04}}\left(x_{0}k_x\right)^2} \right]\\
\tilde{\text{Bu}}\left(k_x,a_2,a_3,k_y\right) &= &x_{0}y_{0} \cdot \delta\left(\left(x_{0} k_x\right)^4 - y_{0}k_y\right) \cdot e^{\text{i}\left(x_{0}k_x\right)^6}e^{\text{i}\frac{a_2}{a_{02}}\left(x_{0}k_x\right)^2}e^{\text{i}\frac{a_3}{a_{03}}\left(x_{0}k_x\right)^3}\\
\tilde{\text{Bu}}\left(k_x,a_2,k_y,a_4\right) &= &x_{0}y_{0} \cdot \delta\left(\left(x_{0} k_x\right)^3 - y_{0}k_y\right) \cdot e^{\text{i}\left(x_{0}k_x\right)^6}e^{\text{i}\frac{a_2}{a_{02}}\left(x_{0}k_x\right)^2}e^{\text{i}\frac{a_4}{a_{04}}\left(x_{0}k_x\right)^4}\\
\tilde{\text{Bu}}\left(k_x,k_y,a_3,a_4\right) &= &x_{0}y_{0} \cdot \delta\left(\left(x_{0} k_x\right)^2 - y_{0}k_y\right) \cdot e^{\text{i}\left(x_{0}k_x\right)^6}e^{\text{i}\frac{a_3}{a_{03}}\left(x_{0}k_x\right)^3}e^{\text{i}\frac{a_4}{a_{04}}\left(x_{0}k_x\right)^4}\\
\end{eqnarray}

In general, the number $N$ of transverse field distributions for each catastrophe can be calculated by considering standard combinatorial methods: $N = \binom{d}{2} = \frac{d!}{2(d-2)!}$, where $d = n-2$ is the co-dimension of the corresponding potential function $P_n$.

\subsection*{Parametrization of the caustics at cross-sections of the swallowtail and butterfly catastrophe}

We performed a parametrization of the caustic surface in dependence of one parameter $u$ and the remaining constant control parameters. The resulting expressions are given in table \ref{tab:SwCaustics} and \ref{tab:BuCaustics}, where the constant control parameters are indicated at the first column. These equations therefore describe cross sections through the three- and four-dimensional control parameter spaces. 

\begin{table}[h]
\centering
\begin{tabular}{|c|c|c|} \hline
$a_1$ & $a_2$ & $a_3$ \\ \hline \hline
$a_2 = 5u^3 - \frac{a_1}{u}$ & $a_1 = 5u^4 - a_2u$ & $a_1 = 15u^4 + 3a_3u^2$ \\ \hline
$a_3 = \frac{a_1}{3u^2} - 5u^2$ & $a_3 = -\frac{a_2}{3u} - \frac{10}{3}u^2$ & $a_2 = -10u^3 - 3a_3u$\\ \hline
\end{tabular}
\caption{Parametrized equations that describe the caustic surface of a swallowtail beam.}
\label{tab:SwCaustics}
\end{table}

\begin{table}[h]
\centering
\begin{tabular}{|c|c|c|}
\hline
$a_1, a_2$ & $a_1, a_3$ & $a_1, a_4$ \\ \hline \hline
$a_3 = 4u^3 - \frac{4a_2}{3u} - \frac{a_1}{u^2}$ & $a_2 = 3u^4 - \frac{3a_1}{4u} - \frac{3}{4}a_3u$ & $a_2 = 2a_4u^2 - \frac{a_1}{u} + 9u^4$ \\ \hline
$a_4 = \frac{a_1}{2u^3} + \frac{a_2}{2u^2} - \frac{9}{2}u^2$ & $a_4 = \frac{a_1}{8u^3} - \frac{3a_3}{8u} - 3u^2$ & $a_3 = \frac{a_1}{3u^2} - \frac{8}{3}a_4u - 8u^3$\\ \hline \hline
$a_2, a_3$ & $a_2, a_4$ & $a_3, a_4$ \\\hline \hline
$a_1 = 4u^5 - a_3u^2 - \frac{4}{3}a_2u$ & $a_1 = 9u^5 + 2a_4u^3 - a_2u$ & $a_1 = 24u^5 + 8a_4u^3 + 3a_3u^2$ \\ \hline
$a_4 = -\frac{a_2}{6u^2} - \frac{a_3}{2u} - \frac{5}{2}u^2$ & $a_3 = -2a_4u - \frac{a_2}{3u} - 5u^3$ & $a_2 = -15u^4 - 6a_4u^2 - 3a_3u$ \\ \hline
\end{tabular}
\caption{Parametrized equations that describe the caustic surface of a butterfly beam.}
\label{tab:BuCaustics}
\end{table}

\subsection*{Experimental Details}
\label{sc:ExpDet}

\begin{figure}[h]
\centering
\includegraphics[width=.7\columnwidth]{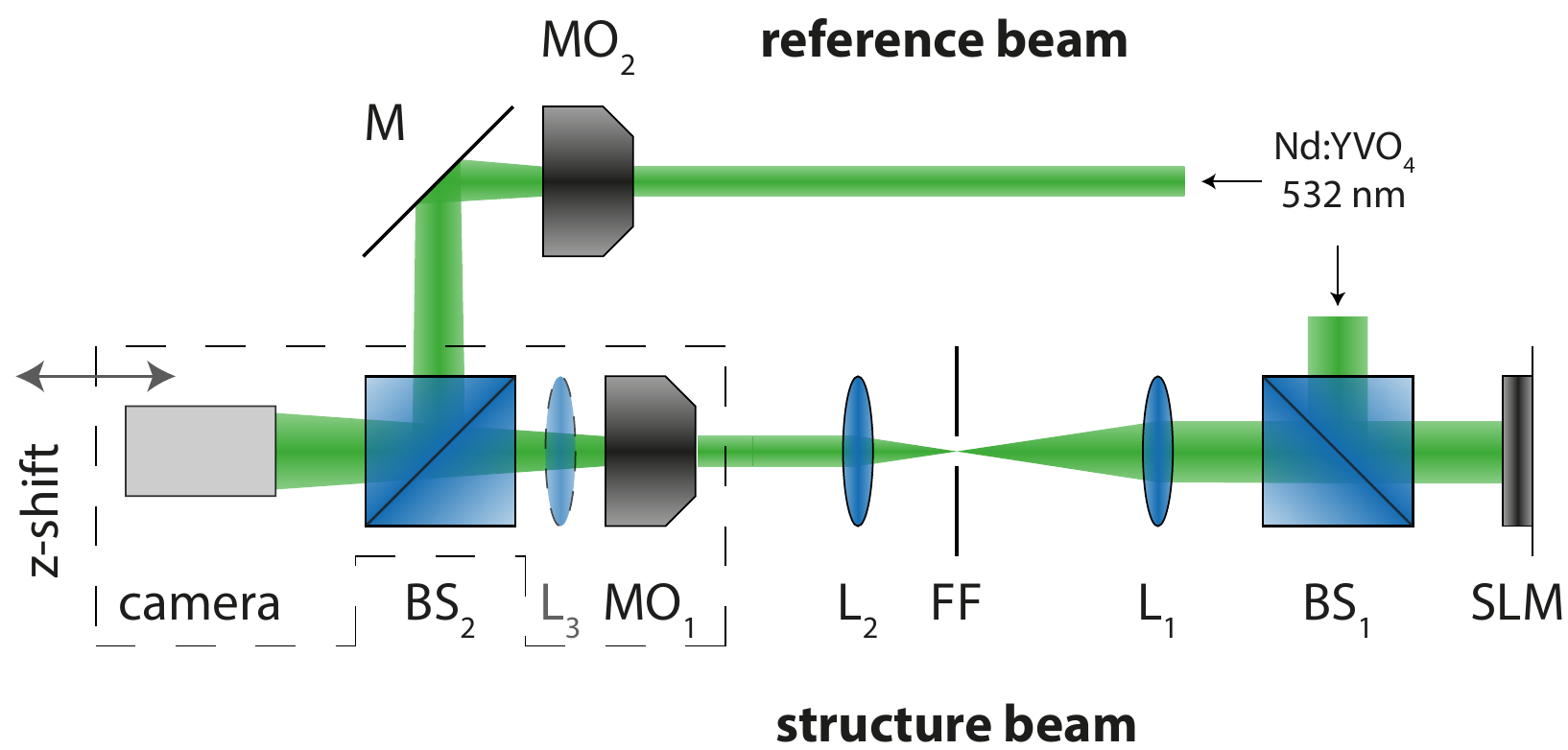}%
\caption{Scheme of experimental setup. BS: beam splitter, FF: Fourier filter, L: lens, M: mirror, MO: microscope objective, SLM: phase-only spatial light modulator.}%
\label{fig:Setup}%
\end{figure}

Fig.~\ref{fig:Setup} shows the experimental setup: An expanded, collimated frequency-doubled Nd:YVO$_4$ laser beam is split up in two parts, which are linearly polarized. One serves as structure beam, whereas the other represents the reference beam. The structure beam is modulated by a HOLOEYE HEO 1080P reflective LCOS phase-only SLM with full HD resolution, which we have applied to simultaneously modulate both amplitude and phase. Therefore, an appropriate Fourier filter (FF) was used. As the imaging system, we mounted a camera and microscope objective with fixed distances on a $z$-shift moving stage that allows us to scan the light field in the longitudinal direction. Measuring the spatial phase distribution is possible by superimposing a tilted reference beam, which can be enabled on demand. By temporarily installing lens L$_3$, the Fourier spectrum becomes accessible. 

\end{document}